How do Bounce Rates vary according to product sold?

Himanshu Sharma

D'Amore-McKim School of Business. Northeastern University

sharma.himanshu4781@gmail.com *

Faculty Advisor: Yakov Bart

**Abstract**

Bounce Rate of different E-commerce websites depends on the different factors based upon the different devices through which traffic share is observed. This research paper focuses on how the type of products sold by different E-commerce websites affects the bounce rate obtained through Mobile/Desktop. It tries to explain the observations which counter the general trend of positive relation between Mobile traffic share and bounce rate and how this is different for the Desktop. To estimate the differences created by the types of products sold by E-commerce websites on the bounce rate according to the data observed for different time, fixed effect model (within group method) is used to determine the difference created by the factors. Along with the effect of the type of products sold by the E-commerce website on bounce rate, the effect of individual website is also compared to verify the results obtained for type of products.

How do Bounce Rates vary according to product sold?

## 1. **Introduction**

Last decade has seen an enormous growth in the traffic shared by Mobile especially since the introduction of smart phones and improvement in the quality of internet all over the world and it seems that the traffic share of Desktop has reduced since then. This pattern is observed for all the kind of mobile websites, including social media, e-commerce websites or any other category of websites. This pattern led to several questions such as "How many people now access 'Facebook' through Desktop?", "How many people have ever accessed "Instagram" through Desktop?" The answer would be very few. Similar kind of questions can be asked for other online websites like "Pin interest", "YouTube", and E-commerce websites as well like "Amazon", "Walmart" etc. This difference in the use of devices to access these websites have motivated E-commerce websites to create mobile friendly websites or application for smart phones. It is now very imperative for E-commerce websites to have proper application or mobile friendly websites to engage their customers. This measure of engagement of customers is observed through Bounce Rate. Bounce rate is very important for E-commerce websites to understand the performance of their page, and helps to understand how well they are able to engage their customers, in other words Bounce Rate gives idea how customers are not able to connect with websites (Lahey, C. , 2020, December 15 and Berman, Barry, 2016, May 24). This is measured when any customer tries to access E-commerce websites and does not perform any actions such as clicking on another link or buying any product on the websites or filling a form. Hence high Bounce Rate indicate poor performance of any E-commerce Websites. And this Bounce Rate is calculated differently for different devices through which traffic is observed. As per the study done by Semrush team on Fintech company (Garanko, J., 2018, August 28), it is observed although the traffic obtained from Mobile was way more than desktop traffic (77%-23%), the bounce rate was way much higher for Mobile users, i.e. the time spent on the website by users was less for Mobile users, and Page/ visit was also less for Mobile users. It seems even though the traffic was much higher but Mobile version was unable to engage the users. There could be multiple reasons for that, may be the mobile version is not comfortable for users to navigate from one page to another, or the mobile app designed for the company is too slow to load that user leave it, or poor quality of internet (Nurminen, J., 2017, April 17). And there other devices also coming in the picture which gives the combination of services of both Mobile and Desktop i.e. Tablet (Enge, E., 2021, March 23). And several research has been done, and different models have been proposed to handle 'Bounce Rate'. Previously in other research papers, factors of online websites like Response Time, Connection Time, Download Time, Page Load Time, Internet connectivity, design of the page and others were considered in explaining the variation in the Bounce Rate (Mittal, Meenakshi, and S. Veena Dhari, 2019). But can there be any other factors which might be able to explain the variation in the Bounce Rate? To understand this it is important to understand the relationship between Bounce Rate and traffic share observed. General observation reveals that as traffic share increases, Bounce Rate also increases especially for the device of Mobile through which constitutes the maximum share of traffic. The study of Semrush.com of Fintech Company (Garanko, J., 2018, August 28) supports this relationship. But can we generalized this pattern observed for Fintech Company (Garanko, J., 2018, August 28) for all E-commerce websites? To answer this question, data for E-commerce websites were observed and this was realized that this relationship between Bounce Rate and traffic share in case of Mobile cannot be generalized. For example for E-commerce websites like cvs.com, adidas.com, and the relationship observed between Bounce Rate and E-commerce websites was different (Figure[1]). It is observed that though the Mobile traffic share was more than the desktop traffic share for this websites, but the Bounce Rate observed through Mobile was less than observed for Desktop. This led to the question that can this relationship between Bounce Rate and Traffic Share depends upon the type of E-commerce websites? Or can this relationship depends on types of products sold by the E-commerce websites? Well it is also important to understand why these questions are important. From customer point of view, lot of discussions has happened on how design of E-commerce websites is mobile friendly to engage the customers in reducing the Bounce Rate. Along with this, size and types of products sell by E-commerce websites also affects the engagement. For example, if a customer wants to buy electronics related products, and want to research more about their specification, then from psychological point of view they should prefer Desktop over Mobile and using Mobile only for the initial research. On the other hand if customer wants to buy Medicines, Jewels, Watches, or other small accessories, then using Mobile would be sufficient to make research about these products. This example led to following hypothesis:

**Ha: In case of Mobile, types of Products sell by the E-commerce websites affects the relationship between Bounce Rate and Traffic Share.**

## 2. **Research Background**

To test the above hypothesis, it is necessary to understand the concept behind time-variant variables and time-invariant variables. The time-varying variables indicate the variables which vary according to time for example days, years etc, on the other hand time-invariant variables are those which are constant from time perspective for example persons, company's name, cities (Cunningham, S.,2021) . This concept is usually used in analyzing the longitudinal data where certain variables vary according to unit over time. This analysis is very important in econometric analysis. And such type of longitudinal data are known as Panel Data. This concept will help us to understand the time invariant effect of the types of products sold by different e-commerce websites on the relationship between time-variant Bounce Rate and time-variant Traffic Share through Mobile. Now it is very important to study the Bounce Rate and Traffic Share through Mobile considering the time effect because it is well known fact that the traffic share obtained though Mobile has increased over time. Hence, considering the time-effect will help us to understand the effect created by the types of products sold by E-commerce websites.

To understand and analyze the effect of time-invariant variables on time-variant variables in the Panel Data, several methods have been prescribed. Following are the methods:

1. Pooled OLS method
2. Fixed/ Random Effect method

To decide which method is best for analyzing the Panel Data, mainly two tests are prescribed namely:

1. Breusch-Pagan test
2. Hausman test

In Breusch-Pagan test, it is tested whether there is an effect of time-invariant variables in the data and hence following null and alternative hypothesis are being considered as follow:

H0: there is no effect of time-invariant variables in the data

HA: there is effect of time-invariant variables in the data

If the p-value is less than 0.05, then H0 is rejected and HA is accepted which will means that there is effect of time-invariant variables in the data.

After this Hausman test is applied to decide if Fixed Effect Model is preferred or Random Effect Model. This test is also pretty straight forward and hence following hypothesis are proposed:

H0: Random Effect is preferred

HA: Fixed Effect is preferred

If the p-value is less than 0.05, then H0 will be rejected, and HA will be accepted i.e. Fixed Effect Model is

## 3. **Data Handling and Conceptual Model**

The data is collected through online Semrush tool for around 1126 different E-commerce websites. The dataset contains the information about the traffic share, bounce rate obtained through Mobile and Desktop for different E-commerce websites. Although there are 24 variables but the research is focused mostly on traffic share obtained through device Mobile and Desktop. The data is collected for each websites from 2017 to 2020 and then aggregated which led to the creation of longitudinal data where the information for each E-commerce websites is recorded over the unit of time from 2017-2020. This aggregation led to 4486 records in the data set. Once the data is collected, the next step is to categorize the different E-commerce websites according to the products sell by different E-commerce websites so that effect of products on the relationship between Bounce Rate and traffic share can be captured. After careful consideration, around 30 categories were created and E-commerce websites were categorized accordingly. To ease the analysis, the abbreviation of 30 categories is created as shown in Table 1.

Now this Panel Data contains the information for each websites from 2017 to 2020. This data is Panel Data from the unit of analysis (Website + Time). But after categorization of these websites, this data no longer remains the Panel Data from the unit of analysis (Category + Time) because there exists multiple values for each category for particular unit of time. Without the Panel Data structure, it is impossible to observe the impact created by the category of each website. To handle this, the data is structured to transform it into Panel data so that for each unit of analysis (Category + Time) there exists a single observation. To transform the data, average values of observation are taken so that for each category and for each time, there exists a single observation. The data is analyzed first for unit of analysis (Category + Time) and then for (Website + Time) and then the results are compared.

To observe the effect through Fixed Effect Model, following model (Figure [2]) is conceptualized. The dependent variable here is Bounce Rate observed, which is affected by the traffic share, but this Bounce Rate is also affected by the type of products sell by the E-commerce websites to observe the effect of type of products. This effect is observed directly and via the traffic shared. And this concept is applied to observe the effect of each individual website on the bounce rate.

## 4. **Estimation Methodology**

Before analyzing the data according to different unit of analysis (Website+Time, Category+Time), two tests Breusch-Pagan test, Hausman test are performed on both original data and transformed data to test if Fixed Effect Models is the best to analyze the data. For Breusch-Pagan test, the p-value obtained for both original data and transformed data are ‘ < 2.2e-16’ and ‘ 0.006303’ respectively and both the values are less than 0.05, and there for null hypothesis is rejected and hence ‘Pooled Ordinary Least Square’ method is not suitable for analyzing the data. After this Hausman Test is performed for both original data and transformed data, and the p-values obtained are ‘< 2.2e-16’ and ‘0.006298’ which are again less than 0.05 and therefore null hypothesis that ‘Random Effect Model’ is preferred is rejected and hence alternative hypothesis “Fixed Effect Model” is preferred (Kurt Schmidheiny, 2021). In Fixed Effect Model, there are several methods out of which ‘within estimator’ method is used to detect effect of category and individual website effect.

For this, first consider the following equation 1

$Y_{it} = \delta D_{it} + u_i + \varepsilon_{it}$ ; $t=1,2,\ldots,T$ ............. (1)

where $Y_{it}$ is the bounce rate observed at time t for time-invariant variable i , $D_{it}$ is the traffic share observed and $u_i$ is the effect of category in transformed data and individual website in original data.

Now average values of traffic share is calculated for all the time period from 2017 to 2020 i.e. t= 1, 2, 3, 4 for each time invariant variable ‘i’ as shown in equation (2):

$Avg(D_i) = (1/T) \sum_{t=1}^{T} (D_{it})$ ...........(2)

Similarly the bounce rate observed through Mobile is calculated for each time invariant for variable ‘i’ with t=1, 2, 3, 4 as shown in equation (3):

$Avg(Y_i) = (1/T) \sum_{t=1}^{T} (Y_{it})$ ................(3)

And similarly calculating average of error term in equation (4):

$Avg(\varepsilon_i) = (1/T) \sum^{T}_{t=1} (\varepsilon_{it})$ ………………(4)

The important thing note here is that average value of $u_i$ for t=1, 2, 3, 4 will be $u_i$ only as it is time-invariant variable.

Now substituting equation (2), (3), (4) in (1):

$Avg(Y_i) = \delta\, Avg(D_i) + u_i + Avg(\varepsilon_i)$ …………(5)

Now time_demeaned values of $D_i$ and $Y_i$ are calculated as follow

$Time_demeaned(D_i) = D_{it} - Avg(D_i)$ ………………(6)

$Time_demeaned(Y_i) = Y_{it} - Avg(Y_i)$ ……………..(7)

$Time_demeaned(\varepsilon_i) = \varepsilon_{it} - Avg(\varepsilon_i)$ ……………..(8)

Equation (1) – Equation (5) led to Equation (9) as follow:

$Y_{it} - Avg(Y_i) = \delta(D_{it} - Avg(D_i)) + (u_i - u_i) + (\varepsilon_{it} - Avg(\varepsilon_i))$ ……………..(9)

In equation, it is observed that $u_i$ value is cancelled out and hence the final equation is obtained by substituting (6), (7) and (8) in (9) as shown below:

$Time_demeaned(Y_i) = \delta(Time_demeaned(D_i)) + Time_demeaned(\varepsilon_i)$ …………(10)

Now this $Time_demeaned(Y_i)$ is regressed on $Time_demeaned(D_i)$ and coefficients are calculated for each time-invariant variable 'i'.

The above explained model is applied separately on both original and transformed data for both Mobile and Desktop.

## 5. **Results and discussion**

There are in total four fixed effect models created for both transformed data and original data. For transformed panel data, the unit of analysis is "Category + Time" and for original data, the unit of analysis is "Website + Time". First model is created on transformed data to capture the effect of category on the relationship between average mobile bounce rate and average mobile share. The p-value (Table 2) obtained for this model is less than <0.001 and R-square value obtained for this model is 94% which indicates the model explains the 94% of variation in the data and the model is good fit. The coefficient obtained for average mobile share is 0.79 which indicates that overall as mobile traffic share increases, mobile bounce rate also increases. Although the coefficients obtained for 30 categories (Table 4) are different, some of them are negative coefficients and some of them are positive coefficients. Negative coefficients obtained for some categories indicates that they have negative contribution towards the relationship of bounce rate and traffic share in case of Mobile. Categories like 'Clothing, Shoes, Jewelry, Watch and other Accessories Retailer", "General" etc have negative coefficients while categories like "Bags and Suitcases Retailers" have positive coefficients. These coefficients (Table 4) indicate that Mobile bounce rate depends on the type of products sold by E-commerce website.

From consumer point of view, it can also be concluded that the products which require minimum research while buying like 'Clothing, Shoes, Jewelry, Watch and other Accessories Retailer", "Eyeglasses, Sunglasses and Contacts Retailer" etc, have negative coefficient which reduces the mobile bounce rate while products which require more research while buying like "Electronics and Technology Retailer", "Fitness and Sports Products and Services retailers" have positive coefficient which indicate that it increases Mobile bounce rate. Similarly, when range of the products sold by any E-commerce website increases, those E-commerce website have negative coefficients which indicate it reduces the bounce rate for such websites. Overall the results obtained for Mobile supports our hypothesis that bounce rate obtained through Mobile depends upon the type of products sell by E-commerce website.

On the other hand, the fixed effect model created for average desktop bounce rate and average desktop traffic share gives different results. The overall p-value (Table 3) is <0.001 and r-square value is 51.2 % which is less than the model obtained for Mobile. This indicates that for desktop, the model explained only 51.2% percent of variation which is very less than 94% obtained for Mobile. The coefficients obtained for different categories(Table 4) are approximately same and all are positive which indicates that in case of desktop all the categories have same effect on the bounce rate. This again supports our hypothesis that the effect of categories on the bounce rate is only valid for Mobile.

Similarly two more fixed effect models created for original data set where unit of analysis is (Website + Time). One model is created to observe the mobile bounce rate and other to observe the desktop bounce rate. For the model created for Mobile (Table 5), the p-value obtained is <0.001 and R-squared value obtained is 82.5% which shows that 82.5% variations are explained by the model which indicate that model is good fit. The coefficients obtained for each websites in case of mobile (Table 7) also supports our hypothesis, for some of the website the coefficient obtained are negative and some are positive. The result is also compared with coefficients obtained in first model for categories. For 635 websites the sign of the coefficients obtained were similar to their corresponding categories, and for rest 491 websites the coefficients obtained were opposite to what we obtained for their corresponding categories.

For model created for Desktop (Table 6), the p-value obtained is <0.001 and R-squared value obtained is just 24% which shows that 24% variations are explained by the model which is way less than the R-square value obtained for Mobile and thus the model is not good fit. The coefficients obtained for websites (Table 7) in case of desktops were all positives which aligns with the result obtained for category specific model created for desktop. Overall we can conclude that incase of desktop there is little effect of both type of websites and type of products sold by E-commerce websites. On the other hand, results shows that that there is larger effect of websites and the types of products sold by E-commerce websites.

## 6. **Managerial implications**

It is clear from the results obtained for the data collected for E-commerce websites, that factors affecting the Bounce Rate obtained through Mobile is different from that of desktop. Although the concept of Bounce Rate is similar for both Desktop and Mobile but data shows that handling of Bounce Rate should be different for Desktop and Mobile. The strategies applied for Desktop to reduce the Bounce Rate might not work for Mobile. Managers need to consider the effect of types of products sold by E-commerce website while applying strategies for reducing the bounce rate. From the data it is observed that for the E-commerce retailers which sells variety of range of products (like general category), and for those E-commerce retailers which sells relatively smaller products or those products which require minimum research while buying have lower bounce rate though Mobile as compared to websites which sells big products or those products which require more research. Along with this, this research is also important for the Startups as well, which will help them to decide which products they should sell by considering the effect of products sold on the bounce rate.

## 7. **Future research**

In this research paper, the main focus is given on the effect of type of products sold by E-commerce websites on the Bounce Rate. Although the result obtained for individual category analysis was compared with the results obtained for individual websites but still tt would be interesting to know the effect of brand value of different E-commerce websites on the bounce rate obtained through Mobile and Desktop. It would also be interesting to know how bounce rate varies for Tablet based on the type of websites and products sold by these E-commerce websites.

## 9. Figures

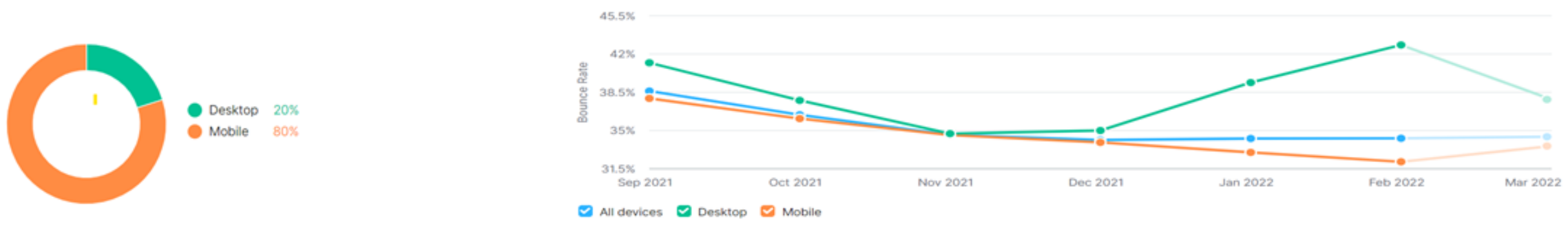

**cvs.com**

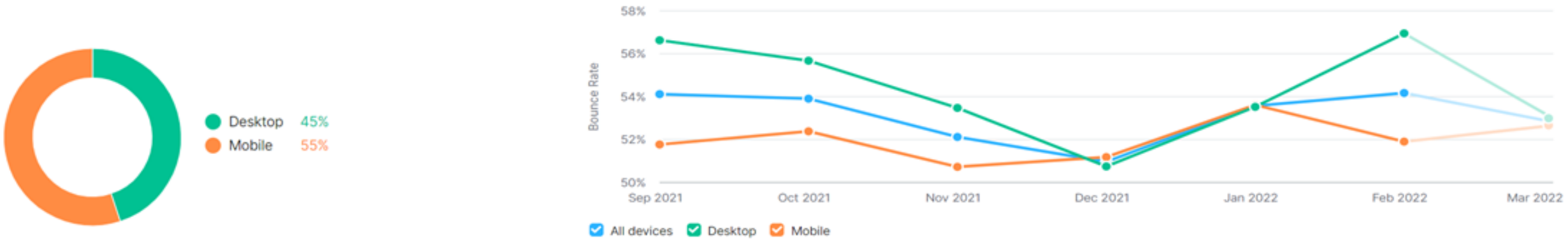

**addidas.com**

Figure 1: Showing the pattern different from general trend for cvs.com and addidas.com

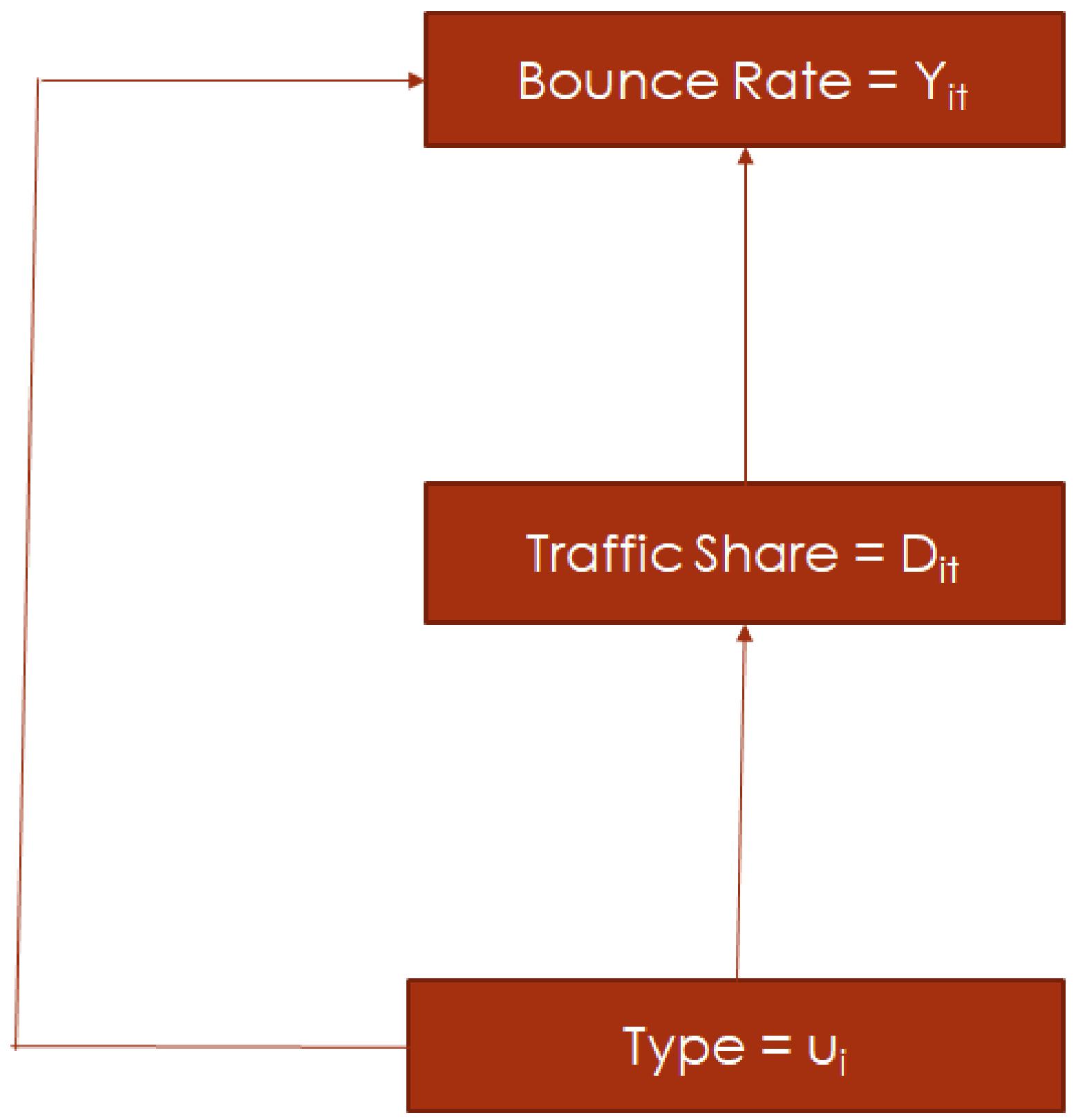

Figure 2: Conceptual Layout displaying the effect of time-invariant on the bounce rate.

## 10. Tables

**Table 1: Categories and Ticker created for the e-commerce websites**

| Ticker | Category |
|---|---|
| **AAR** | Autoparts and Automobile Retailer |
| **ACFR** | Arts, Crafts and Fabric Retailers |
| **BPR** | Baby Products Retailers |
| **BSR** | Bags and Suitcases Retailer |
| **CSJWAR** | Clothing, Shoes, Jewelery, Watch and Accessories Retailer |
| **CVPR** | Cigarette & Vape Prodcuts Retailer |
| **ESCR** | Eyeglasses, Sunglasses and Contacts Retailer |
| **ETR** | Electronics and Technology Retailer |
| **FBGR** | Foods,Beverages and Groceries Retailer |
| **FSPSR** | Fitness & Sports Products and Services Retailer |
| **G** | General |
| **GPR** | Gardening Products Retailer |
| **GTR** | Gifts and Toys Retailer |
| **HICPR** | Home Improvement and Cleaning Products Retailer |
| **HSPPR** | Health Services and Pharamceutical Products Retailers |
| **KTER** | Kitchen tools and Equipments Retailer |
| **MEEPR** | Media, Education and Entertainment Products Retailer |
| **MEP** | Marketing and E-commerce Platform |
| **MPR** | Metal Products Retailer |
| **MPSR** | Music Products and Services Retailer |
| **MS** | Mail Services |
| **ORMR** | Outdoor Recreation Merchandise Retailer |
| **OSSR** | Office and Stationary Supplies Retailer |
| **PCBPR** | Personal Care and Beauty Products Retailer |
| **PDS** | Packaging And Distribution Services |
| **PLPR** | Printing and Labels Products Retailer |
| **PPEMSR** | Party Products and Event Management Services Retailer |
| **PPR** | Pet Products Retailer |
| **PVER** | Photo and Video Equipment Retailer |
| **WITER** | Weapons, Industrial Tools and Equipment Retailer |

**Table 2: Showing the Fixed Effect Model Category Specific Result for average_mobile_bounce_rate**

| | **Dependent variable** | | |
|---|---|---|---|
| *Predictors* | *Estimates* | *CI* | *p* |
| df$average_mobile_share | 0.79 | 0.75 – 0.83 | **<0.001** |
| Observations | 120 | | |
| $R^2$ / $R^2$ adjusted | 0.940 / 0.920 | | |

**Table 3: Showing the Fixed Effect Model Category Specific Result for average_desktop_bounce_rate**

| | Dependent variable | | |
|---|---|---|---|
| *Predictors* | *Estimates* | *CI* | *p* |
| df$average_desktop_share | 0.18 | 0.14 – 0.22 | **<0.001** |
| Observations | 120 | | |
| $R^2$ / $R^2$ adjusted | 0.512 / 0.348 | | |

**Table 4: Showing the result for coefficient of different categories for Mobile and Desktop**

| Ticker | Category | For Mobile | For Desktop |
|---|---|---|---|
| **AAR** | Autoparts and Automobile Retailer | **-0.0850641** | 0.294982 |
| **ACFR** | Arts, Crafts and Fabric Retailers | **-0.0189456** | 0.250623 |
| **BPR** | Baby Products Retailers | 0.0197489 | 0.284849 |
| **BSR** | Bags and Suitcases Retailer | 0.0805583 | 0.265251 |
| **CSJWAR** | Clothing, Shoes, Jewelery, Watch and Accessories Retailer | **-0.0284684** | 0.228568 |
| **CVPR** | Cigarette & Vape Prodcuts Retailer | **-0.0928377** | 0.315462 |
| **ESCR** | Eyeglasses, Sunglasses and Contacts Retailer | **-0.0250915** | 0.223294 |
| **ETR** | Electronics and Technology Retailer | 0.0719491 | 0.313977 |
| **FBGR** | Foods,Beverages and Groceries Retailer | **-0.0018116** | 0.278785 |
| **FSPSR** | Fitness & Sports Products and Services Retailer | 0.0229327 | 0.271882 |
| **G** | General | **-0.0396396** | 0.285387 |
| **GPR** | Gardening Products Retailer | **-0.0117521** | 0.394577 |
| **GTR** | Gifts and Toys Retailer | **-0.0277441** | 0.262765 |
| **HICPR** | Home Improvement and Cleaning Products Retailer | **-0.004536** | 0.281803 |
| **HSPPR** | Health Services and Pharamceutical Products Retailers | **-0.0139404** | 0.325029 |
| **KTER** | Kitchen tools and Equipments Retailer | 0.0240129 | 0.299464 |
| **MEEPR** | Media, Education and Entertainment Products Retailer | 0.042154 | 0.249406 |
| **MEP** | Marketing and E-commerce Platform | 0.1483329 | 0.046402 |
| **MPR** | Metal Products Retailer | **-0.0015401** | 0.270729 |
| **MPSR** | Music Products and Services Retailer | 0.0391399 | 0.313499 |
| **MS** | Mail Services | 0.2417515 | 0.383415 |
| **ORMR** | Outdoor Recreation Merchandise Retailer | **-0.0766455** | 0.293999 |
| **OSSR** | Office and Stationary Supplies Retailer | 0.1058442 | 0.309996 |
| **PCBPR** | Personal Care and Beauty Products Retailer | **-0.0608098** | 0.256735 |
| **PDS** | Packaging And Distribution Services | 0.0600703 | 0.223107 |
| **PLPR** | Printing and Labels Products Retailer | 0.1757746 | 0.286945 |
| **PPEMSR** | Party Products and Event Management Services Retailer | **-0.0537645** | 0.276109 |
| **PPR** | Pet Products Retailer | **-0.0816716** | 0.281684 |
| **PVER** | Photo and Video Equipment Retailer | 0.0890465 | 0.237417 |

**Table 5: Showing the Fixed Effect Model Website Specific Result for mobile_bounce_rate**

| | **Dependent variable** | | |
|---|---|---|---|
| *Predictors* | *Estimates* | *CI* | *p* |
| bounceplm$mobile.share | 0.70 | 0.69 – 0.71 | **<0.001** |
| Observations | 4480 | | |
| $R^2$ / $R^2$ adjusted | 0.825 / 0.766 | | |

**Table 6: Showing the Fixed Effect Model Website Specific Result for desktop_bounce_rate**

| | **Dependent variable** | | |
|---|---|---|---|
| *Predictors* | *Estimates* | *CI* | *p* |
| bounceplm$desktop.share | 0.17 | 0.16 – 0.18 | **<0.001** |
| Observations | 4472 | | |
| $R^2$ / $R^2$ adjusted | 0.240 / -0.015 | | |

**Table 7: Showing the result for coefficient of different websites for Mobile and Desktop**

| Column1 | Column2 | Column3 |
|---|---|---|
| Website | Website specific Mobile coefficients | Website specific Desktop coefficients |
| 1000bulbs.com | 0.077908101 | 0.347694436 |
| 123stores.com | -0.034609246 | 0.516496057 |
| 13deals.com | 0.187851858 | 0.417938994 |
| 1800contacts.com | 0.022875566 | 0.35738294 |
| 1800flowers.com | -0.088006344 | 0.168204064 |
| 1800gunsandammo.com | 0.162016323 | 0.47786426 |
| 1800petmeds.com | -0.005084135 | 0.283221512 |
| 1sale.com | 0.038673791 | 0.442371792 |
| 42photo.com | 0.094093594 | 0.441599297 |
| 47stphoto.com | 0.043518723 | 0.379367118 |
| 4inkjets.com | 0.182690831 | 0.439630651 |
| 4wd.com | 0.020207966 | 0.366767782 |
| 511tactical.com | -0.033438082 | 0.258508656 |
| 6dollarshirts.com | 0.016583466 | 0.296899266 |
| abercrombie.com | 0.028100537 | 0.126202925 |
| abt.com | 0.129422045 | 0.371992733 |
| accessories4less.com | 0.050087003 | 0.240606339 |
| aclens.com | 0.15492433 | 0.48122479 |
| acmetools.com | 0.002651115 | 0.363139637 |
| acnestudios.com | 0.267182265 | 0.077663696 |
| acwholesalers.com | 0.068315248 | 0.408162607 |
| adidas.com | -0.016290704 | 0.266465425 |
| adorama.com | 0.174388133 | 0.366691901 |
| adoreme.com | -0.18931406 | 0.4339164 |
| ae.com | -0.020346216 | 0.129880965 |
| aedsuperstore.com | 0.237021987 | 0.332400313 |
| aeropostale.com | -0.082880445 | 0.116289789 |
| agacistore.com | -0.054164679 | 0.227936436 |
| agentprovocateur.com | 0.060056541 | 0.226323256 |
| agrisupply.com | -0.011159018 | 0.332108101 |
| aimsurplus.com | -0.059057405 | 0.311072621 |
| aircompressorsdirect.com | 0.048155224 | 0.369033669 |
| airsoftgi.com | -0.012314825 | 0.208975393 |
| ajmadison.com | 0.19984301 | 0.389962185 |
| aldoshoes.com | 0.013197303 | 0.380766028 |
| alexandani.com | -0.058756361 | 0.198835608 |
| alibris.com | 0.054566235 | 0.294115267 |
| aliengearholsters.com | 0.005026084 | 0.337621063 |
| allbirds.com | 0.044536141 | 0.179371645 |
| alliedelec.com | 0.273673758 | 0.344740066 |
| allparts.com | 0.118566379 | 0.203423439 |
| allsaints.com | 0.191514108 | 0.119489499 |
| alphabroder.com | 0.168502135 | -0.008181512 |
| amainhobbies.com | -0.031705373 | 0.200336995 |
| amazon.com | 0.035744106 | 0.135717235 |
| americanapparel.com | 0.125275466 | 0.137878179 |
| americanfloormats.com | 0.111878586 | 0.402920416 |
| americangirl.com | 0.039327537 | 0.120361425 |
| americanmuscle.com | -0.091603011 | 0.25513952 |
| americanmusical.com | 0.000570788 | 0.270203666 |
| americansignaturefurniture.com | -0.007294289 | 0.202343592 |
| americanstandard-us.com | 0.054151789 | 0.248135504 |
| amerimark.com | -0.187725978 | 0.210933218 |
| amiclubwear.com | -0.00389257 | 0.241162908 |
| amway.com | -0.082403594 | 0.121925582 |
| andersenwindows.com | -0.03361956 | 0.214044232 |
| angara.com | 0.075092677 | 0.437148408 |
| anker.com | 0.090038066 | 0.495017127 |
| anntaylor.com | -0.074411166 | 0.152875176 |
| antonline.com | 0.22354445 | 0.165820971 |
| apmex.com | -0.015196898 | 0.270913499 |
| apple.com | 0.281573226 | 0.385008994 |
| appliancepartspros.com | -0.008119459 | 0.315283957 |
| appliancesconnection.com | 0.138640485 | 0.409395207 |
| aquasana.com | 0.043107086 | 0.363894664 |
| aritzia.com | 0.167797142 | 0.051008964 |
| armaniexchange.com | 0.046123451 | 0.148948896 |
| art.com | 0.090661779 | 0.301105168 |
| article.com | 0.118760606 | 0.1903477 |
| artvan.com | -0.09965331 | 0.290924476 |
| ashford.com | -0.032720729 | 0.355551994 |
| ashleystewart.com | -0.173696473 | 0.278444492 |
| asics.com | -0.037254638 | 0.215588619 |
| asos.com | 0.052451644 | 0.114048927 |
| audio-technica.com | 0.243912758 | 0.305341595 |
| austinbazaar.com | 0.257899599 | 0.385221643 |
| autoaccessoriesgarage.com | -0.002107164 | 0.360215512 |
| autonomous.ai | 0.242553265 | 0.522350436 |
| autopartswarehouse.com | -0.01960995 | 0.39136335 |
| autoplicity.com | 0.099606197 | 0.572536007 |
| autozone.com | -0.129914945 | 0.28395898 |
| avenue.com | -0.140674426 | 0.203600498 |
| avery.com | 0.169857843 | 0.140297221 |
| avon.com | -0.141624638 | 0.273429543 |
| awaytravel.com | 0.164372845 | 0.402420669 |
| babyhaven.com | 0.121561528 | 0.408212707 |
| backcountry.com | 0.044372348 | 0.377958627 |
| balsamhill.com | 0.097922209 | 0.295227664 |
| barcodesinc.com | 0.245917819 | 0.330260949 |
| bareminerals.com | -0.065020362 | 0.186928734 |
| barenecessities.com | 0.026303688 | 0.264292245 |
| bargainstation.com | 0.448231595 | 0.577261339 |
| barkbox.com | -0.079573998 | 0.271230144 |
| barneys.com | 0.182033847 | 0.255909225 |
| bassettfurniture.com | 0.029581395 | 0.180407636 |
| batesfootwear.com | 0.012093429 | 0.186723898 |
| baublebar.com | 0.097945251 | 0.234381999 |
| bcbg.com | 0.045785491 | 0.189315669 |
| beachbody.com | 0.306477791 | 0.343693022 |
| beachcamera.com | 0.056783979 | 0.404703006 |
| beallsflorida.com | -0.123118202 | 0.298717018 |
| beautyencounter.com | 0.058387613 | 0.436582718 |
| beautylish.com | -0.035955292 | 0.412100168 |
| bebe.com | -0.004483434 | 0.198733278 |
| beckertime.com | 0.092203552 | 0.578709323 |
| bedbathandbeyond.com | 0.000453755 | 0.314044424 |
| belk.com | -0.077994177 | 0.209779264 |
| belkin.com | 0.120194305 | 0.37645819 |
| bellacor.com | 0.124543218 | 0.344159614 |
| benchmade.com | -0.017711698 | 0.190229862 |
| benq.us | 0.409858348 | 0.415111754 |
| bespokepost.com | -0.056590165 | 0.456962403 |
| bestbuy.com | -0.018417709 | 0.22244239 |
| betabrand.com | 0.026077352 | 0.526429028 |
| betterworldbooks.com | 0.038890146 | 0.287504848 |
| beyondtherack.com | 0.093187889 | 0.218226408 |
| bhcosmetics.com | -0.094645486 | 0.220823759 |
| bhfo.com | -0.060595128 | 0.29172452 |
| bhphotovideo.com | 0.225691718 | 0.340422004 |
| big5sportinggoods.com | -0.147649787 | 0.183454228 |
| bigbadtoystore.com | -0.081002378 | 0.191564539 |
| bigceramicstore.com | 0.168054564 | 0.414263945 |
| biglots.com | -0.196098698 | 0.172532543 |
| bikebandit.com | 0.002613029 | 0.322788014 |
| billabong.com | 0.040916686 | 0.141767039 |
| birchbox.com | -0.017280093 | 0.494683861 |
| bisonoffice.com | 0.050718895 | 0.437208059 |
| bissell.com | 0.001623113 | 0.282514839 |
| bizchair.com | 0.140895228 | 0.327540949 |
| bjs.com | -0.10690057 | 0.316515812 |
| bkstr.com | -0.030255899 | 0.245586157 |
| blackanddecker.com | -0.011227508 | 0.318723426 |
| blackdiamondequipment.com | 0.189814381 | 0.139837511 |
| blackforestdecor.com | -0.006113296 | 0.242795051 |
| blacklapel.com | 0.308929605 | 0.511610312 |
| bladehq.com | -0.095671344 | 0.229908549 |
| blissworld.com | 0.078579364 | 0.433881753 |
| blucoil.com | 0.437350868 | 0.457951353 |
| blueapron.com | 0.063183084 | 0.183743508 |
| bluefly.com | 0.137217987 | 0.297495953 |
| bluenile.com | 0.042068774 | 0.278026959 |
| bn.com | 0.219762078 | 0.563826491 |
| boats.net | -0.024154164 | 0.313166875 |
| bobswatches.com | 0.118755888 | 0.306579305 |
| bodenusa.com | -0.012374542 | 0.164765527 |
| bodyartforms.com | -0.04833346 | 0.205424818 |
| bodycandy.com | -0.094718591 | 0.25968417 |
| bollandbranch.com | 0.026260858 | 0.257467172 |
| bombas.com | -0.117033389 | 0.204644005 |
| bonton.com | -0.074922234 | 0.240088043 |
| bookbyte.com | 0.202941362 | 0.271092858 |
| bootbarn.com | -0.066640882 | 0.172090596 |
| bosch-home.com | 0.109909126 | 0.283814913 |
| boscovs.com | -0.180045768 | 0.237300165 |
| bose.com | 0.110725749 | 0.28466168 |
| bostonproper.com | -0.098042179 | 0.264558817 |
| bouclair.com | 0.15415659 | 0.227467951 |
| boxed.com | -0.013012829 | 0.301021773 |
| boxycharm.com | -0.080563568 | 0.205926294 |
| brambleberry.com | -0.024639125 | 0.199478717 |
| brandless.com | -0.048213436 | 0.459886074 |
| brandsmartusa.com | -0.069836994 | 0.493268832 |
| brantano.com.mx | -0.017145125 | 0.237292702 |
| brentwoodhome.com | 0.122507191 | 0.302091687 |
| brevilleusa.com | 0.234536275 | 0.228420454 |
| brighton.com | -0.137030329 | 0.152416727 |
| brilliantearth.com | 0.060473336 | 0.2670531 |
| brooklinen.com | 0.151932334 | 0.18531117 |
| brooksbrothers.com | 0.144331455 | 0.15029945 |
| brooksrunning.com | 0.011031152 | 0.222342457 |
| brookstone.com | 0.067307767 | 0.289315937 |
| brownells.com | -0.029185508 | 0.360990087 |
| btosports.com | 0.055105521 | 0.321692613 |
| buckle.com | -0.048725274 | 0.223675554 |
| budsgunshop.com | -0.121875834 | 0.283532123 |
| buggiesunlimited.com | -0.271301284 | 0.321753099 |
| build.com | -0.070055861 | 0.355889844 |
| buildasign.com | 0.154812797 | 0.225685322 |
| builddirect.com | 0.160352028 | 0.359804301 |
| bulbamerica.com | 0.096876419 | 0.383538535 |
| bulbs.com | 0.148479563 | 0.406175251 |
| bulkapothecary.com | 0.040857764 | 0.259752822 |
| bulkreefsupply.com | 0.029699711 | 0.233841097 |
| bungiestore.com | 0.015451065 | 0.252548112 |
| burberry.com | 0.049171289 | 0.160403486 |
| burlington.com | -0.075327253 | 0.432937276 |
| burpee.com | 0.066584213 | 0.371715703 |
| burton.co.uk | 0.301509171 | 0.592704371 |
| burton.com | 0.224896167 | 0.182595986 |
| buyautoparts.com | -0.090640856 | 0.398629557 |
| buycostumes.com | 0.267300526 | 0.370858703 |
| buydig.com | 0.107478035 | 0.367490557 |
| buyonlinenow.com | 0.300835369 | 0.439959695 |
| c21stores.com | -0.029851365 | 0.156799313 |
| cabelas.com | -0.07201944 | 0.252011633 |
| cabinetparts.com | 0.050247832 | 0.320497361 |
| cabinets.com | 0.08872833 | 0.251869948 |
| cafepress.com | 0.098812761 | 0.406473242 |
| calendars.com | -0.053396599 | 0.339516079 |
| calvinklein.com | 0.140094295 | 0.312510513 |
| campingworld.com | -0.085386606 | 0.282431189 |
| campsaver.com | 0.116612291 | 0.366750838 |
| canadagoose.com | 0.234645038 | 0.138965406 |
| canadiantire.ca | 0.011367758 | 0.244381794 |
| cardsdirect.com | 0.276153403 | 0.427066705 |
| carhartt.com | -0.020134434 | 0.175949341 |
| carid.com | -0.062479624 | 0.361257606 |
| carters.com | -0.125676426 | 0.142356224 |
| casetify.com | 0.222283633 | 0.155400249 |
| casper.com | 0.058605336 | 0.24051776 |
| catalog.usmint.gov | -0.002632907 | 0.170790738 |
| catofashions.com | -0.219921405 | 0.101988227 |
| cavenders.com | -0.129539199 | 0.154377744 |
| cduniverse.com | 0.031171526 | 0.391451378 |
| cdw.com | 0.337905205 | 0.271094142 |
| cellularoutfitter.com | -0.126697419 | 0.347767856 |
| chanel.com | 0.124170936 | 0.206974451 |
| chapters.indigo.ca | 0.294671626 | 0.239740716 |
| charbroil.com | 0.017499093 | 0.444012191 |
| charlesandcolvard.com | 0.012229434 | 0.204702894 |
| charlotterusse.com | 0.062197334 | 0.103860529 |
| charmingcharlie.com | -0.051980719 | 0.256531211 |
| cheaperthandirt.com | -0.061658617 | 0.334511395 |
| chefknivestogo.com | 0.016219793 | 0.208433486 |
| chefsplate.com | 0.252556092 | 0.506676571 |
| chewy.com | -0.112578122 | 0.212407093 |
| chicagomusicexchange.com | 0.148566642 | 0.431920433 |
| chicos.com | -0.12415395 | 0.209564825 |
| childrensplace.com | -0.155731134 | 0.187619973 |
| christianbook.com | -0.039566978 | 0.284510641 |
| christmascentral.com | 0.127439779 | 0.347572237 |
| christopherandbanks.com | -0.174755869 | 0.187430047 |
| cigarsinternational.com | -0.137723244 | 0.226945374 |
| citizenwatch.com | -0.065764778 | 0.291794235 |
| claires.com | -0.022666975 | 0.198311942 |
| clarksusa.com | -0.099291065 | 0.151158824 |
| claroshop.com | 0.044097832 | 0.351227395 |
| classicfirearms.com | -0.134624716 | 0.386650737 |
| clickbank.com | 0.158645597 | 0.054303369 |
| cmp.callawaygolf.com | -0.019603016 | 0.186141533 |
| coach.com | 0.034797221 | 0.210571014 |
| colehaan.com | 0.052633609 | 0.201134069 |
| colemanfurniture.com | 0.127730467 | 0.238259001 |
| collectionsetc.com | -0.224492304 | 0.241810943 |
| columbia.com | -0.009790757 | 0.204768983 |
| compacc.com | -0.06920788 | 0.406150668 |
| connection.com | 0.330840329 | 0.348926544 |
| containerstore.com | 0.03905639 | 0.187175212 |
| cookiesbydesign.com | -0.065378684 | 0.324274635 |
| cookieskids.com | -0.077636023 | 0.262883392 |
| coolibar.com | -0.037818505 | 0.212960484 |
| coolstuffinc.com | -0.047507147 | 0.231864075 |
| corelle.com | -0.059528859 | 0.219317496 |
| cosstores.com | 0.147695502 | 0.081895346 |
| costco.com | -0.060321029 | 0.15577415 |
| cottonon.com | 0.052520454 | 0.074388986 |
| cpooutlets.com | -0.023250918 | 0.348430096 |
| crashplan.com | 0.433407617 | 0.255875953 |
| crateandbarrel.com | 0.121824571 | 0.176812938 |
| creekstonefarms.com | 0.088139227 | 0.285901497 |
| crocs.com | -0.002721584 | 0.198201563 |
| crownandcaliber.com | 0.041163409 | 0.302080302 |
| crucial.com | 0.223450158 | 0.361941075 |
| crutchfield.com | -0.02264988 | 0.290543881 |
| ctshirts.com | 0.019115846 | 0.124800973 |
| customink.com | 0.127759557 | 0.160565839 |
| cutco.com | -0.063683268 | 0.266186942 |
| cutleryandmore.com | -0.003049683 | 0.25681519 |
| cvs.com | -0.164794305 | 0.304226687 |
| cymax.com | 0.074493969 | 0.390011759 |
| dafiti.com.br | -0.105216476 | 0.311015489 |
| dailyburn.com | 0.219753739 | 0.501400708 |
| dailylook.com | -0.01733014 | 0.276534162 |
| danscomp.com | 0.003094116 | 0.145529542 |
| databazaar.com | 0.410214883 | 0.461695698 |
| davidsbridal.com | -0.045002298 | 0.163125101 |
| dbrand.com | 0.096337623 | 0.170345356 |
| deepdiscount.com | 0.024376323 | 0.264847285 |
| dell.com | 0.142556253 | 0.265599909 |
| delmarfans.com | 0.191916853 | 0.443005008 |
| denon.com | 0.100951288 | 0.341588198 |
| destinationmaternity.com | -0.012640235 | 0.385867665 |
| destinationxl.com | -0.059753769 | 0.372719525 |
| dexclusive.com | 0.261954262 | 0.42295324 |
| dia.com | -0.13022821 | 0.245067642 |
| diamondcandles.com | -0.144800022 | 0.213915243 |
| diamondnexus.com | -0.031558859 | 0.303637645 |
| dickblick.com | 0.040197757 | 0.288159867 |
| dickssportinggoods.com | -0.063087506 | 0.189894965 |
| diesel.com | 0.103368722 | 0.206607784 |
| digikey.com | 0.321745146 | 0.214078255 |
| dillards.com | -0.007632715 | 0.257789602 |
| directron.com | 0.221201857 | 0.418738243 |
| directvapor.com | -0.093111161 | 0.244165978 |
| discountbodyparts.com | -0.047608905 | 0.358436587 |
| discountdance.com | 0.014833905 | 0.105222825 |
| discountfilters.com | -0.067973801 | 0.328730999 |
| discountramps.com | -0.024676316 | 0.393870331 |
| discounttire.com | -0.08248094 | 0.193403907 |
| discoverystore.com | 0 | 0.553422309 |
| dog.com | 0.058763489 | 0.489822841 |
| doitbest.com | -0.035050046 | 0.329060045 |
| dollarshaveclub.com | -0.097669827 | 0.283923963 |
| dollartree.com | -0.096586862 | 0.265091299 |
| dollskill.com | 0.018666303 | 0.13074208 |
| domyown.com | 0.05971429 | 0.48527276 |
| dooney.com | -0.140053442 | 0.193870189 |
| doversaddlery.com | -0.005037787 | 0.156923033 |
| drjays.com | -0.034600592 | 0.181115276 |
| dsw.com | -0.089909498 | 0.422509029 |
| dtlr.com | 0.117567533 | 0.297981264 |
| duluthtrading.com | -0.114451319 | 0.187403721 |
| dungarees.com | -0.112597107 | 0.286435962 |
| dvf.com | 0.133456481 | 0.281764841 |
| dynamiteclothing.com | 0.174818798 | 0.38140157 |
| dyson.com | 0.108784038 | 0.234306193 |
| eaze.com | -0.069155171 | 0.488460604 |
| ec-club.panasonic.jp | -0.042668672 | 0.014779174 |
| ecampus.com | 0.103010796 | 0.157816754 |
| ecstuning.com | 0.017029189 | 0.269988754 |
| eddiebauer.com | -0.068882886 | 0.159475467 |
| ediblearrangements.com | -0.110405551 | 0.142361757 |
| eero.com | 0.212011101 | 0.502033393 |
| efaucets.com | 0.079399738 | 0.461741074 |
| eforcity.com | 0.117560105 | 0.567633539 |
| eileenfisher.com | 0.003581755 | 0.174114816 |
| elderly.com | -0.018324893 | 0.286998215 |
| elektra.com.mx | -0.071017073 | 0.253924551 |
| elementvape.com | -0.097384217 | 0.224250718 |
| elfcosmetics.com | -0.065854847 | 0.177555333 |
| elitefts.com | 0.091498556 | 0.383772058 |
| eloquii.com | -0.088594309 | 0.20743622 |
| elpalaciodehierro.com | 0.109502284 | 0.249995899 |
| emusic.com | 0.280627242 | 0.438749832 |
| en-us.sennheiser.com | 0.239516717 | 0.182580427 |
| endclothing.com | 0.29564115 | 0.325908777 |
| entertainmentearth.com | 0.035568905 | 0.376143463 |
| ereplacementparts.com | -0.068237602 | 0.328079644 |
| ericdress.com | -0.040839942 | 0.331594779 |
| esalerugs.com | -0.051672579 | 0.241306357 |
| esalon.com | -0.057942903 | 0.31939507 |
| eshakti.com | -0.07033982 | 0.182758533 |
| essential.com | 0.007506162 | 0.469619778 |
| essentialhardware.com | 0.308246366 | 0.376960669 |
| esteelauder.com | -0.008114716 | 0.2597039 |
| ethanallen.com | 0.050853496 | 0.126365745 |
| etsy.com | 0.05630598 | 0.173052906 |
| evacuumstore.com | 0.048615675 | 0.354141879 |
| everlane.com | 0.130222942 | 0.43517863 |
| everythingkitchens.com | 0.076353874 | 0.365079991 |
| evine.com | -0.156166988 | 0.218027064 |
| evo.com | 0.220539196 | 0.261734504 |
| explodingkittens.com | 0.123720693 | 0.484212817 |
| express.com | -0.008440787 | 0.135055122 |
| fabfitfun.com | -0.168683877 | 0.210188219 |
| factoryauthorizedoutlet.com | 0.049402661 | 0.452049923 |
| factoryoutletstore.com | -0.09262901 | 0.358254842 |
| famousfootwear.com | -0.108554778 | 0.170994513 |
| famsa.com | 0.04379671 | 0.175490038 |
| fanatics.com | -0.032376815 | 0.193359472 |
| farmandfleet.com | -0.066655522 | 0.281514786 |
| fashionmia.com | 0.049893975 | 0.321789801 |
| fashionnova.com | -0.093207043 | 0.107711939 |
| fast-growing-trees.com | 0.038482364 | 0.343703262 |
| fatbraintoys.com | 0.03551466 | 0.324714271 |
| fathead.com | 0.050246059 | 0.298450531 |
| faucetdepot.com | 0.329000246 | 0.293840106 |
| fender.com | 0.097522195 | 0.285976604 |
| fergusonshowrooms.com | 0.031762521 | 0.237186488 |
| fineartamerica.com | 0.119300937 | 0.335497825 |
| fingerhut.com | -0.164623092 | 0.111501972 |
| finishline.com | 0.052712962 | 0.199938984 |
| firemountaingems.com | -0.036412555 | 0.265753811 |
| fishusa.com | -0.080539163 | 0.457728092 |
| fitbit.com | -0.038975929 | 0.235419717 |
| flexshopper.com | -0.118833361 | 0.143362363 |
| flightclub.com | 0.156060375 | 0.254669249 |
| flooranddecor.com | -0.030990338 | 0.21989012 |
| focuscamera.com | 0.238086123 | 0.392211829 |
| foodsaver.com | -0.051443142 | 0.306365494 |
| footlocker.com | 0.082723852 | 0.291699096 |
| footsmart.com | -0.082167174 | 0.383721761 |
| forever21.com | 0.004976903 | 0.099317822 |
| forhims.com | 0.003896105 | 0.545658707 |
| foryourlegs.com | -0.077659934 | 0.306650137 |
| fosmon.com | 0.103217952 | 0.3979444 |
| fossil.com | -0.002586226 | 0.163459357 |
| fragrancenet.com | -0.101697222 | 0.198750721 |
| fragrantjewels.com | -0.147493137 | 0.189893576 |
| francescas.com | 0.022734711 | 0.05773649 |
| freshdirect.com | 0.131039827 | 0.057811415 |
| freshwatersystems.com | 0.234575129 | 0.396898362 |
| fromyouflowers.com | -0.087318601 | 0.215159248 |
| frys.com | -0.045334478 | 0.289078354 |
| ftd.com | 0.018415216 | 0.361012337 |
| fullbeauty.com | -0.182482197 | 0.201423204 |
| fullcompass.com | 0.238255966 | 0.325264513 |
| fullsource.com | 0.044333125 | 0.326310377 |
| furla.com | 0.037041212 | 0.132869768 |
| furniture.com | 0.123962865 | 0.424647673 |
| furniturerow.com | -0.032103445 | 0.199945684 |
| fye.com | -0.046621735 | 0.299695833 |
| gaiam.com | 0.248528086 | 0.462807144 |
| gamefly.com | -0.032648082 | 0.481008812 |
| gamestop.com | -0.04777405 | 0.192201231 |
| gap.com | -0.108845093 | 0.140090753 |
| gardeners.com | 0.076092187 | 0.448980068 |
| gardensalive.com | 0.142789618 | 0.585230492 |
| gearbest.com | 0.196838045 | 0.433638663 |
| gearhead.com | 0.160516525 | 0.484644273 |
| gemselect.com | 0.161471471 | 0.468716244 |
| gemvara.com | 0.034672967 | 0.273457497 |
| gerbergear.com | -0.080211461 | 0.236642789 |
| getfpv.com | 0.051600629 | 0.241787104 |
| ghostbed.com | 0.135169554 | 0.260023091 |
| gifttree.com | 0.155540198 | 0.280569016 |
| glassesusa.com | -0.013606177 | 0.219609267 |
| globalgolf.com | -0.114243131 | 0.187269386 |
| globalindustrial.com | 0.184245841 | 0.36338443 |
| glossier.com | 0.149159699 | 0.162661931 |
| gnc.com | -0.008125559 | 0.257367403 |
| godiva.com | -0.037274877 | 0.274322182 |
| goldsilver.com | 0.223852084 | 0.370078445 |
| golfballs.com | -0.044264139 | 0.24547092 |
| gourmetgiftbaskets.com | 0.150588274 | 0.341584572 |
| govbergwatches.com | 0.147911333 | 0.361167764 |
| govx.com | -0.135431894 | 0.25726168 |
| grabagun.com | -0.140653022 | 0.333215879 |
| grainger.com | 0.034179166 | 0.334028637 |
| grasscity.com | 0.111519197 | 0.48736787 |
| graze.com | -0.162405257 | 0.243535534 |
| greenchef.com | -0.027747125 | 0.407470088 |
| greenhousemegastore.com | 0.047063526 | 0.264589125 |
| grizzly.com | -0.047469478 | 0.253019293 |
| groupon.com | -0.003429758 | 0.300539505 |
| grove.co | -0.15280578 | 0.3231936 |
| gruntstyle.com | -0.09297734 | 0.208009175 |
| guitarcenter.com | 0.013371442 | 0.229569724 |
| gymboree.com | 0.032890175 | 0.274656564 |
| hallmark.com | 0.009932987 | 0.360522965 |
| halloweencostumes.com | 0.203763473 | 0.325013023 |
| hammacher.com | -0.075402611 | 0.439177621 |
| hanes.com | -0.07515823 | 0.224549874 |
| hannaandersson.com | 0.066240965 | 0.185780697 |
| harborfreight.com | -0.154983526 | 0.29733428 |
| harley-davidson.com | -0.098269094 | 0.254468202 |
| harmankardon.com | 0.149740055 | 0.374728406 |
| harristeeter.com | -0.179961115 | 0.170968957 |
| harrys.com | -0.036606626 | 0.199880363 |
| harvestright.com | 0.128039539 | 0.256676052 |
| hasbropulse.com | 0.117091376 | 0.193555257 |
| hasbrotoyshop.com | 0.081493827 | 0.320031644 |
| hbx.com | 0.303879377 | 0.237967936 |
| healingcrystals.com | 0.193812522 | 0.461766238 |
| healthydirections.com | -0.023482907 | 0.596558479 |
| heartratemonitorsusa.com | -0.016111915 | 0.375598093 |
| heels.com | 0.192072283 | 0.585263809 |
| heirloomroses.com | -0.023753552 | 0.341961547 |
| helixsleep.com | 0.004938492 | 0.179162737 |
| hellofresh.com | -0.020712727 | 0.186077061 |
| herimports.com | 0.086469911 | 0.212467695 |
| hermanmiller.com | 0.245039352 | 0.227816694 |
| herroom.com | -0.00539209 | 0.289542637 |
| hhgregg.com | 0.023222159 | 0.373021096 |
| hibbett.com | 0.043707157 | 0.21683766 |
| hickoryfarms.com | -0.111588164 | 0.36595635 |
| hipsandcurves.com | -0.055009016 | 0.224661116 |
| hm.com | 0.020021614 | 0.099861201 |
| hobbylobby.com | -0.079635937 | 0.173722038 |
| homechef.com | -0.023072838 | 0.223388312 |
| homedepot.com | -0.037741568 | 0.219348635 |
| honest.com | 0.029675274 | 0.261705093 |
| hoover.com | -0.009806324 | 0.353151817 |
| hottopic.com | -0.040710672 | 0.198746058 |
| hubblecontacts.com | 0.031512328 | 0.140904593 |
| humankinetics.com | 0.301717409 | 0.350022339 |
| hy-vee.com | -0.124533197 | 0.105505324 |
| ibuyofficesupply.com | 0.335526668 | 0.463154924 |
| ice.com | 0.135033107 | 0.339615696 |
| icetrends.com | -0.093686983 | 0.133809236 |
| iherb.com | 0.042521407 | 0.284083192 |
| ikea.com | -0.058908758 | 0.116222988 |
| improvementscatalog.com | 0.200527302 | 0.525550134 |
| indochino.com | 0.138213157 | 0.11473066 |
| infowarsstore.com | -0.062857766 | 0.191070434 |
| instawares.com | 0.371563604 | 0.622962076 |
| ipsy.com | -0.15218878 | 0.177381294 |
| jabra.com | 0.161618719 | 0.330777493 |
| jacksonandperkins.com | -0.082577753 | 0.271107433 |
| jackssmallengines.com | -0.083773864 | 0.299684429 |
| jackthreads.com | 0.145403514 | 0.113065167 |
| jameco.com | 0.204371774 | 0.384283526 |
| jamesavery.com | -0.110995217 | 0.270019525 |
| jansport.com | 0.041692976 | 0.244705913 |
| jaybirdsport.com | 0.107598195 | 0.303777692 |
| jcp.com | 0.146218957 | 0.447832112 |
| jcpenney.com | -0.151175891 | 0.204951956 |
| jcrew.com | 0.006539329 | 0.164579422 |
| jdsports.co.uk | 0.209101021 | 0.319185395 |
| jefferspet.com | 0.010481065 | 0.291861736 |
| jeffreestarcosmetics.com | -0.06824746 | 0.064775568 |
| jegs.com | 0.004537446 | 0.354048947 |
| jerrysartarama.com | -0.009050125 | 0.258010444 |
| jetpens.com | 0.032260815 | 0.182177284 |
| jewlr.com | -0.038097923 | 0.497994167 |
| jgsales.com | -0.04917512 | 0.292921346 |
| jhilburn.com | 0.153558684 | 0.120766852 |
| jimmyjazz.com | -0.002536586 | 0.265623969 |
| jjbuckley.com | 0.303045521 | 0.307640444 |
| jjill.com | -0.126381818 | 0.155636481 |
| jmbullion.com | -0.001011252 | 0.359848906 |
| joann.com | -0.061541784 | 0.199922083 |
| jockey.com | -0.025082397 | 0.234651978 |
| joesnewbalanceoutlet.com | -0.124826564 | 0.244665038 |
| johnsonfitness.com | 0.125544456 | 0.3181721 |
| jomashop.com | -0.018138944 | 0.342757348 |
| joybird.com | 0.089417295 | 0.2839375 |
| joyus.com | 0.100628436 | 0.440988637 |
| jpcycles.com | -0.06015253 | 0.267830293 |
| jtv.com | -0.13106328 | 0.252507022 |
| justfab.com | -0.18613002 | 0.1193779 |
| juul.com | 0.014592901 | 0.315291732 |
| jwpepper.com | 0.186530233 | 0.120072301 |
| kaiusaltd.com | 0.03925168 | 0.219398449 |
| karmaloop.com | 0.130623831 | 0.203389203 |
| kay.com | -0.095963775 | 0.191471193 |
| keenfootwear.com | -0.061412597 | 0.176345167 |
| keh.com | 0.167809776 | 0.167965244 |
| kellycodetectors.com | -0.022839676 | 0.3303131 |
| kendrascott.com | -0.018082219 | 0.130546399 |
| kennethcole.com | 0.045900145 | 0.232543358 |
| keurig.com | -0.05274529 | 0.247708792 |
| kikocosmetics.com | 0.127795393 | 0.142945455 |
| kingarthurflour.com | 0.058035952 | 0.508532738 |
| kirklands.com | -0.17412831 | 0.189309039 |
| kitchenaid.com | 0.085674565 | 0.271606792 |
| kiwico.com | 0.011777239 | 0.325837662 |
| klipsch.com | 0.056888981 | 0.385091749 |
| klwines.com | 0.104771026 | 0.345470165 |
| knifecenter.com | -0.02146235 | 0.337875116 |
| knoll.com | 0.30719191 | 0.168452935 |
| kohls.com | -0.120497833 | 0.16238549 |
| kpoptown.com | 0.122987405 | 0.099271477 |
| kraftmusic.com | 0.117698392 | 0.336589215 |
| kroger.com | -0.127110973 | 0.273967071 |
| kuiu.com | -0.075225365 | 0.185044726 |
| kultofathena.com | -0.10977999 | 0.149626163 |
| kyliecosmetics.com | 0.015630179 | 0.147581189 |
| la-z-boy.com | -0.090393123 | 0.156086379 |
| lacoste.com | 0.001764646 | 0.207787359 |
| lafayette148ny.com | 0.10606992 | 0.171888366 |
| lakeshorelearning.com | 0.032910543 | 0.241466018 |
| lakeside.com | -0.148320603 | 0.209636923 |
| lampsplus.com | 0.046046549 | 0.257483118 |
| lancasterarchery.com | -0.019641219 | 0.276210774 |
| landsend.com | -0.087500485 | 0.180887371 |
| lapolicegear.com | -0.113950366 | 0.287658892 |
| leatherman.com | -0.003992969 | 0.218268388 |
| leesa.com | 0.076920099 | 0.269547738 |
| leevalley.com | 0.019093955 | 0.324728493 |
| lehmans.com | -0.041622092 | 0.38296658 |
| lenovo.com | 0.059591167 | 0.319374617 |
| lesliespool.com | -0.023283595 | 0.263982853 |
| letote.com | 0.034987576 | 0.155088234 |
| levenger.com | -0.046911262 | 0.24529384 |
| levi.com | 0.027163061 | 0.175165739 |
| lexmark.com | 0.226694784 | 0.255601267 |
| lids.com | -0.039871739 | 0.178728296 |
| lifetime.com | 0.045686916 | 0.311526973 |
| lifeway.com | 0.070903856 | 0.238057948 |
| lightinthebox.com | -0.078771649 | 0.340981694 |
| linentablecloth.com | -0.073545786 | 0.238690759 |
| liverpool.com.mx | -0.004699229 | 0.226031589 |
| livingsocial.com | -0.020386422 | 0.355839372 |
| livingspaces.com | 0.015818007 | 0.207943463 |
| llbean.com | -0.121582032 | 0.199159688 |
| loblaws.ca | 0.227926759 | 0.408309699 |
| lollywollydoodle.com | 0.00679407 | 0.19113858 |
| londondrugs.com | 0.137712229 | 0.32621222 |
| longislandwatch.com | -0.050698109 | 0.340764224 |
| lootcrate.com | -0.096213013 | 0.220388169 |
| lorealparisusa.com | 0.030654483 | 0.388582816 |
| lorextechnology.com | 0.026220815 | 0.29884418 |
| lovelyskin.com | 0.054373586 | 0.36417096 |
| lovesac.com | -0.011691165 | 0.166153634 |
| lowes.com | -0.084099743 | 0.161273722 |
| luckybrand.com | -0.019049763 | 0.131895987 |
| luckygunner.com | 0.012533707 | 0.413317434 |
| luisaviaroma.com | 0.299902011 | 0.234535462 |
| lulus.com | 0.040279567 | 0.133703532 |
| lulzbot.com | 0.334228018 | 0.291199055 |
| lumberliquidators.com | -0.047533324 | 0.245083395 |
| lunacycle.com | 0.08557186 | 0.232123915 |
| luxedecor.com | 0.151137016 | 0.337961773 |
| macys.com | -0.029306309 | 0.182115778 |
| mancrates.com | -0.091030457 | 0.333176062 |
| mantelsdirect.com | 0.047047296 | 0.33697049 |
| marineengine.com | 0.006717801 | 0.392677404 |
| marti.mx | 0.267540949 | 0.049830761 |
| matterhackers.com | 0.231145571 | 0.476365152 |
| mattressfirm.com | -0.072884644 | 0.262727276 |
| mavistire.com | -0.136598228 | 0.222497206 |
| mec.ca | 0.143621211 | 0.333207841 |
| meijer.com | -0.152389826 | 0.176888826 |
| mellanni.com | 0.042248919 | 0.192213259 |
| menards.com | -0.162461453 | 0.159867117 |
| menswearhouse.com | -0.040032487 | 0.250053686 |
| merchbar.com | 0.006037544 | 0.395508022 |
| meundies.com | -0.041026309 | 0.130339471 |
| mgemi.com | 0.120557585 | 0.187931353 |
| michaelkors.com | -0.015688547 | 0.262281906 |
| michaels.com | -0.102407651 | 0.168700587 |
| michiganbulb.com | -0.131466943 | 0.284098246 |
| microcenter.com | -0.018060895 | 0.133012334 |
| microsoft.com | 0.189247028 | 0.406262588 |
| midwayusa.com | -0.062707343 | 0.385295775 |
| mileskimball.com | -0.11659752 | 0.341399185 |
| mint.ca | 0.067161842 | 0.238403652 |
| minted.com | 0.196303675 | 0.133642944 |
| mlbshop.com | 0.012640045 | 0.145700596 |
| mmlafleur.com | 0.138785397 | 0.212535353 |
| modaoperandi.com | 0.163307177 | 0.373793074 |
| monkeysports.com | 0.109159005 | 0.521638877 |
| monoprice.com | 0.127694901 | 0.202074525 |
| monrovia.com | 0.103057099 | 0.476497285 |
| montblanc.com | 0.115155926 | 0.250261456 |
| morphe.com | 0.224609768 | 0.333450008 |
| motionrc.com | -0.03951736 | 0.262157154 |
| motosport.com | -0.062486135 | 0.292149351 |
| mouser.com | 0.281286332 | 0.277917623 |
| moviemars.com | 0.125677558 | 0.035302018 |
| mscdirect.com | 0.115317724 | 0.213876301 |
| munchery.com | 0.386092759 | 0.35108703 |
| musicgoround.com | 0.025443117 | 0.454856909 |
| musiciansfriend.com | 0.037519575 | 0.43801355 |
| musicnotes.com | 0.24317432 | 0.309288653 |
| mvmtwatches.com | -0.002789139 | 0.160935404 |
| mybobs.com | -0.12937688 | 0.091177393 |
| mygoods.com | 0.073346443 | 0.419022091 |
| myotcstore.com | 0.016769207 | 0.318604157 |
| mypillow.com | -0.091728451 | 0.207416434 |
| myvaporstore.com | -0.02891646 | 0.240031097 |
| nakedwines.com | -0.070796943 | 0.160434977 |
| napaonline.com | -0.058862574 | 0.20092717 |
| nastygal.com | 0.146285992 | 0.115697677 |
| natchezss.com | -0.093707216 | 0.366153538 |
| nationalbusinessfurniture.com | 0.299742903 | 0.412264871 |
| naturebox.com | 0.072786873 | 0.365086925 |
| nautilus.com | 0.270902043 | 0.365964463 |
| nectarsleep.com | 0.079493789 | 0.611656639 |
| needsupply.com | 0.265376194 | 0.209018164 |
| neimanmarcus.com | 0.146433767 | 0.232082621 |
| nespresso.com | 0.051817298 | 0.120606003 |
| net-a-porter.com | 0.292233493 | 0.290171856 |
| neutrogena.com | 0.108137935 | 0.366871849 |
| newark.com | 0.217968607 | 0.445787394 |
| newbalance.com | 0.004257189 | 0.210022798 |
| newegg.com | 0.041931092 | 0.238596872 |
| nflshop.com | -0.006897209 | 0.170595833 |
| nfm.com | -0.073511018 | 0.086773437 |
| nike.com | 0.085377125 | 0.146893204 |
| nikonusa.com | 0.144077521 | 0.375116233 |
| ninewest.com | -0.145004227 | 0.227750887 |
| nixon.com | 0.043343676 | 0.197630484 |
| nomnomnow.com | 0.078530675 | 0.333732342 |
| nordictrack.com | 0.125962251 | 0.344175298 |
| nordstrom.com | 0.013048677 | 0.286752234 |
| northerntool.com | -0.084609568 | 0.38425668 |
| northshorecare.com | 0.021406974 | 0.251362108 |
| nutrisystem.com | -0.118866099 | 0.273047711 |
| nuts.com | 0.082555188 | 0.33163629 |
| nyandcompany.com | -0.189428834 | 0.149117821 |
| officedepot.com | 0.038898179 | 0.182278122 |
| officesupply.com | 0.098400362 | 0.388760023 |
| officeworld.com | 0.121800817 | 0.485260677 |
| omahasteaks.com | -0.090403277 | 0.283252339 |
| omega.com | 0.297744332 | 0.335309946 |
| onecall.com | 0.116356051 | 0.527253608 |
| oneill.com | -0.021893426 | 0.186473746 |
| onepeloton.com | 0.094400578 | 0.275777843 |
| onkyousa.com | 0.04264575 | 0.361371643 |
| onlinelabels.com | 0.218489982 | 0.252665089 |
| onlinestores.com | 0.153281633 | 0.449989405 |
| onpurple.com | 0.244799711 | 0.464615195 |
| opgi.com | -0.022568156 | 0.232582815 |
| opticsplanet.com | 0.020659324 | 0.399413133 |
| oreillyauto.com | -0.123637065 | 0.199642664 |
| orientaltrading.com | -0.046292092 | 0.218469049 |
| origamiowl.com | -0.165568183 | 0.084878172 |
| orvis.com | -0.030494964 | 0.264527947 |
| ospreypacks.com | 0.234794166 | 0.53549005 |
| oster.com | 0.055208108 | 0.288324597 |
| otterbox.com | 0.03470799 | 0.170107882 |
| ourpamperedhome.com | 0.120764474 | 0.334944036 |
| outdoorresearch.com | 0.051737325 | 0.189673592 |
| overnightprints.com | 0.342570085 | 0.140179782 |
| overstock.com | -0.004919065 | 0.304711525 |
| overtons.com | -0.070152247 | 0.33340892 |
| oxo.com | 0.121227483 | 0.369436532 |
| pacsun.com | 0.186590188 | 0.109098687 |
| painfulpleasures.com | 0.058801387 | 0.375238656 |
| paintball-online.com | 0.06881507 | 0.366969811 |
| palmbeachjewelry.com | -0.192685008 | 0.203359505 |
| palmettostatearmory.com | -0.129430978 | 0.286870111 |
| pamperedchef.com | 0.006867883 | 0.102923523 |
| pampers.com | 0.094713139 | 0.333869826 |
| papyrusonline.com | 0.016819658 | 0.249218409 |
| parachutehome.com | 0.167294072 | 0.245701129 |
| parts-express.com | 0.028882939 | 0.302245976 |
| parts.andersenwindows.com | -0.073732462 | 0.130165917 |
| partselect.com | -0.011140266 | 0.347535179 |
| partstree.com | -0.110382757 | 0.235280937 |
| partycity.com | -0.04812272 | 0.202567542 |
| patagonia.com | 0.113873959 | 0.127690077 |
| paulaschoice.com | 0.07461561 | 0.376855064 |
| paulfredrick.com | -0.066684449 | 0.27049658 |
| payless.com | -0.041669795 | 0.330701793 |
| pcconnection.com | 0.34764337 | 0.573255236 |
| pcliquidations.com | 0.102871333 | 0.445219384 |
| pcrichard.com | 0.006880209 | 0.263803899 |
| peapod.com | -0.070026259 | 0.377565816 |
| pepboys.com | -0.111705084 | 0.186831016 |
| performancebike.com | 0.071647393 | 0.2040915 |
| perryellis.com | 0.062474653 | 0.298006507 |
| peruvianconnection.com | 0.040982076 | 0.265283908 |
| petco.com | -0.079327225 | 0.22343664 |
| petflow.com | -0.032941056 | 0.182054903 |
| pgatoursuperstore.com | -0.06761807 | 0.187653203 |
| pgshop.com | 0.040553602 | 0.392455362 |
| pharmapacks.com | 0.103410147 | 0.372787045 |
| pier1.com | -0.099541223 | 0.169392088 |
| plantdelights.com | 0.137043734 | 0.463971159 |
| plantronics.com | 0.137764778 | 0.325201347 |
| play.google.com | -0.118137018 | 0.151919185 |
| plumbingsupply.com | 0.161385359 | 0.56530616 |
| plushbeds.com | 0.053645969 | 0.506110373 |
| poppin.com | 0.21618468 | 0.306175536 |
| popsockets.com | -0.046816983 | 0.177069443 |
| potpourrigift.com | -0.138383318 | 0.451012689 |
| powells.com | 0.108270472 | 0.343942205 |
| prada.com | 0.163341846 | 0.128416717 |
| prepsportswear.com | 0.055276999 | 0.374891141 |
| pricechopper.com | -0.112628304 | 0.225232677 |
| primaryarms.com | -0.07328436 | 0.516168779 |
| primestyle.com | 0.036600665 | 0.512110408 |
| princessauto.com | 0.111201318 | 0.257616197 |
| printingforless.com | 0.408916198 | 0.471977644 |
| printm3d.com | 0.137002441 | 0.289736631 |
| prodirectsoccer.com | 0.119681796 | 0.123304751 |
| promgirl.com | 0.175524739 | 0.136685437 |
| propercloth.com | 0.21106632 | 0.28863002 |
| propertyroom.com | -0.015380604 | 0.196592558 |
| provantage.com | 0.323463366 | 0.377545003 |
| providentmetals.com | 0.007691376 | 0.295354191 |
| pssl.com | 0.123888167 | 0.294755503 |
| publix.com | -0.197409777 | 0.13545303 |
| puravidabracelets.com | -0.000772131 | 0.109188502 |
| purchasingpower.com | -0.174181876 | 0.067372397 |
| pureformulas.com | 0.110857177 | 0.349326599 |
| puritan.com | -0.028688909 | 0.288929388 |
| purplecarrot.com | 0.077949688 | 0.206829902 |
| qalo.com | -0.043521589 | 0.222243157 |
| qualitylogoproducts.com | 0.272544468 | 0.48659921 |
| qvc.com | -0.105663279 | 0.176358465 |
| rackroomshoes.com | -0.088323897 | 0.152534579 |
| radioshack.com | 0.080902563 | 0.325575319 |
| radpowerbikes.com | -0.013666919 | 0.276047561 |
| rag-bone.com | 0.242006421 | 0.088664619 |
| rainbowshops.com | -0.227543922 | 0.208180237 |
| ralphlauren.com | 0.057482282 | 0.153593091 |
| ray-ban.com | -0.014224641 | 0.162045981 |
| raymourflanigan.com | -0.094569633 | 0.159512335 |
| rcwilley.com | -0.074452385 | 0.142895866 |
| realtruck.com | -0.045013173 | 0.405716672 |
| rebeccaminkoff.com | 0.05083019 | 0.239545792 |
| rebelcircus.com | 0.020125139 | 0.405650535 |
| reddressboutique.com | -0.007839038 | 0.27212396 |
| redwingheritage.com | 0.050996384 | 0.200636902 |
| rei.com | 0.042606752 | 0.227147373 |
| reitmans.com | 0.006515264 | 0.358419957 |
| relaxtheback.com | 0.119872615 | 0.418695438 |
| renttherunway.com | -0.01241146 | 0.108139252 |
| repairclinic.com | 0.011046347 | 0.401635527 |
| replacements.com | -0.025475905 | 0.249393019 |
| restorationhardware.com | 0.128659357 | 0.140020579 |
| revolve.com | 0.119481521 | 0.181996901 |
| revzilla.com | -0.018006094 | 0.286829194 |
| rickis.com | 0.028127391 | 0.250994311 |
| ritani.com | 0.188504194 | 0.430397958 |
| rkguns.com | -0.152274684 | 0.344478381 |
| roadrunnersports.com | 0.098927956 | 0.243268142 |
| robotshop.com | 0.275077541 | 0.383957824 |
| rockauto.com | -0.090525537 | 0.353841932 |
| rockbottomgolf.com | -0.126971079 | 0.21811837 |
| rockler.com | -0.027331536 | 0.34343838 |
| romwe.com | -0.011770133 | 0.178614496 |
| roomandboard.com | 0.150920767 | 0.176261951 |
| roomstogo.com | -0.090401258 | 0.262209096 |
| roots.com | 0.158318093 | 0.123803659 |
| ross-simons.com | -0.089209101 | 0.315519112 |
| ross-tech.com | 0.081649299 | 0.366122653 |
| rothys.com | -0.065679676 | 0.350995638 |
| rubbermaid.com | 0.08917844 | 0.252849604 |
| rue21.com | -0.095560622 | 0.121025695 |
| ruelala.com | -0.128662403 | 0.163866137 |
| rugs-direct.com | 0.047350768 | 0.217750497 |
| rugstudio.com | 0.093272 | 0.277715022 |
| rugsusa.com | -0.00952523 | 0.192513722 |
| runningwarehouse.com | 0.004149639 | 0.148598471 |
| ruralking.com | -0.156869925 | 0.312525557 |
| rushordertees.com | 0.161551108 | 0.117539475 |
| ruvilla.com | 0.155409995 | 0.665757314 |
| saatvamattress.com | 0.023781868 | 0.321421329 |
| saberforge.com | -0.133918326 | 0.109877119 |
| saddlebackleather.com | 0.133894943 | 0.322948855 |
| sallybeauty.com | -0.099743008 | 0.25745845 |
| samash.com | -0.088134606 | 0.309591305 |
| samys.com | 0.209528316 | 0.211653762 |
| sanrio.com | 0.132437725 | 0.270064234 |
| saxxunderwear.com | 0.014019211 | 0.14856315 |
| scentsy.com | -0.092820163 | 0.098398585 |
| scheels.com | -0.072732659 | 0.167775977 |
| schiit.com | 0.07284328 | 0.204525772 |
| scholastic.com | 0.243098943 | 0.081773975 |
| schooloutfitters.com | 0.186191707 | 0.305831449 |
| scrubsandbeyond.com | -0.007664432 | 0.254295594 |
| sdbullion.com | 0.039689293 | 0.220426707 |
| sdwheel.com | 0.143570802 | 0.234809368 |
| sears.com | -0.058276913 | 0.324068152 |
| searsoutlet.com | -0.09937038 | 0.409989868 |
| selectblinds.com | 0.049513971 | 0.173623246 |
| sephora.com | -0.004759262 | 0.164211548 |
| serta.com | 0.11533963 | 0.263952674 |
| sfplanet.com | 0.222544268 | 0.313399704 |
| sgammo.com | -0.005534687 | 0.25694709 |
| sharkrobot.com | 0.12525946 | 0.279556472 |
| sharperimage.com | 0.063119402 | 0.40671305 |
| sheetmusicplus.com | 0.224157513 | 0.443222653 |
| shethinx.com | 0.10685041 | 0.18757235 |
| shi.com | 0.342262073 | 0.203658905 |
| shindigz.com | 0.116101096 | 0.389521426 |
| shinesty.com | 0.119263803 | 0.145364562 |
| shinola.com | 0.183708895 | 0.212355934 |
| shoecarnival.com | -0.138180908 | 0.214521312 |
| shoemall.com | -0.016449917 | 0.326305059 |
| shoesforcrews.com | -0.115828013 | 0.232736465 |
| shoeshow.com | -0.009280077 | 0.471813889 |
| shoeshowmega.com | -0.198150584 | 0.170423813 |
| shoesofprey.com | 0.172258693 | 0.363781429 |
| shop.1sale.com | -0.110928094 | 0.104321465 |
| shop.advanceautoparts.com | -0.056444369 | 0.246994618 |
| shop.americangreetings.com | -0.238196722 | -0.01530847 |
| shop.com | 0.107303793 | 0.296487312 |
| shop.concept2.com | 0.027922961 | 0.136332443 |
| shop.cricut.com | -0.221475608 | 0.115744353 |
| shop.gonoodle.com | -0.236611746 | -5.36E-05 |
| shop.gopro.com | -0.027442457 | 0.068460845 |
| shop.guess.com | 0.062803646 | 0.13818594 |
| shop.hodinkee.com | 0.12652798 | 0.168704561 |
| shop.lego.com | -0.065226094 | 0.121797493 |
| shop.lululemon.com | 0.188130446 | 0.090400167 |
| shop.nationalgeographic.com | 0.046002519 | 0.148541004 |
| shop.nhl.com | -0.019422064 | 0.123127164 |
| shop.panasonic.com | 0.15337974 | 0.304148664 |
| shop.safeway.com | -0.205083303 | 0.104164066 |
| shop.samsonite.com | 0.085262821 | 0.210033187 |
| shop.surfboard.com | 0.017480846 | 0.263525835 |
| shop.turtlebeach.com | -0.355388161 | 0.061581556 |
| shop.usa.canon.com | 0.128783934 | 0.200880884 |
| shop.wornandwound.com | -0.239356655 | -0.016548788 |
| shop.wwe.com | -0.145944257 | 0.194774132 |
| shop.xerox.com | 0.180122778 | 0.242995059 |
| shopandroid.com | 0.058233006 | 0.488992176 |
| shopdisney.com | -0.038833949 | 0.154541325 |
| shopjimmy.com | 0.069194542 | 0.209546492 |
| shopladder.com | 0.101206829 | 0.460266444 |
| shoplet.com | 0.155138035 | 0.229371903 |
| shopmyexchange.com | -0.134931569 | 0.138358308 |
| shopperschoice.com | 0.312292639 | 0.306171613 |
| shoprite.com | -0.145981367 | 0.088875651 |
| shoptiques.com | 0.202879551 | 0.355790719 |
| shopworldkitchen.com | 0.193247476 | 0.346398946 |
| shutterfly.com | 0.058210824 | 0.119069203 |
| sierratradingpost.com | 0.071408509 | 0.475715447 |
| signaturehardware.com | 0.10601866 | 0.306887135 |
| signs.com | 0.267664842 | 0.333643311 |
| simons.ca | -0.001682704 | 0.217576672 |
| skagen.com | -0.014501281 | 0.219769836 |
| skechers.com | -0.102985748 | 0.229215236 |
| skinit.com | 0.000604617 | 0.233806619 |
| skis.com | 0.158748447 | 0.375393878 |
| skullcandy.com | 0.06619469 | 0.192806902 |
| skymall.com | 0.025098488 | 0.349564983 |
| sleepnumber.com | 0.016437088 | 0.302610023 |
| smarthome.com | 0.1530543 | 0.480620056 |
| smartpakequine.com | 0.074533954 | 0.168897805 |
| smartsign.com | 0.151459499 | 0.32154572 |
| smartwool.com | 0.167759935 | 0.198022978 |
| smiledirectclub.com | 0.04035999 | 0.229087764 |
| smokecartel.com | 0.091538621 | 0.421870755 |
| soccer.com | 0.070344011 | 0.191775951 |
| sofamania.com | 0.040647964 | 0.259857357 |
| softchoice.com | 0.487871655 | 0.271939097 |
| softsurroundings.com | -0.06000612 | 0.209538868 |
| solesociety.com | 0.004109929 | 0.266293311 |
| solidsignal.com | 0.068111788 | 0.460818247 |
| sonicelectronix.com | -0.052145126 | 0.367079931 |
| sonos.com | 0.112572719 | 0.329993442 |
| soriana.com | 0.024738481 | 0.226787031 |
| soundstrue.com | 0.101422347 | 0.364435942 |
| spanx.com | 0.044817959 | 0.275558937 |
| sparkfun.com | 0.322695879 | 0.376723232 |
| speedwaymotors.com | 0.04993881 | 0.492650276 |
| spencersonline.com | -0.098248956 | 0.216920517 |
| sportbiketrackgear.com | 0.032772644 | 0.318038266 |
| sportique.com | 0.204282746 | 0.388808305 |
| sportsmans.com | -0.012956913 | 0.439509292 |
| sportsmanswarehouse.com | 0.21296248 | 0.333525705 |
| spreadshirt.com | 0.1284682 | 0.501036494 |
| spud.ca | 0.428144173 | 0.181043327 |
| ssactivewear.com | 0.27531346 | 0.044998689 |
| ssense.com | 0.234563853 | 0.427325269 |
| stacksocial.com | 0.24014882 | 0.302376599 |
| stage.com | -0.16108833 | 0.318403949 |
| stamps.com | 0.281017353 | 0.388557203 |
| staples.com | 0.021592129 | 0.249739551 |
| steelseries.com | 0.221703008 | 0.268755114 |
| steinmart.com | -0.149731464 | 0.258026978 |
| stevemadden.com | 0.070139542 | 0.121723839 |
| stevespanglerscience.com | 0.362545687 | 0.407139797 |
| stewmac.com | 0.026959236 | 0.280325771 |
| stitchfix.com | 0.056499524 | 0.004909269 |
| stopzilla.com | -0.092120152 | 0.35012575 |
| store.americanapparel.net | -0.021354081 | 0.203433637 |
| store.creekstonefarms.com | -0.349213121 | 0.055207422 |
| store.cyberweld.com | -0.235814083 | 0.232120634 |
| store.discovery.com | 0.011913531 | 0.249112388 |
| store.gearpatrol.com | -0.137862796 | 0.119606587 |
| store.google.com | -0.11524712 | 0.185102674 |
| store.hp.com | 0.06195574 | 0.234952806 |
| store.irobot.com | -0.043907288 | 0.292262306 |
| store.nba.com | 0.146561949 | 0.109590591 |
| store.onkyousa.com | -0.15474334 | 0.015403953 |
| store.printm3d.com | 0.082592564 | 0.12810901 |
| store.schoolspecialty.com | 0.120745103 | 0.062510541 |
| store.snapon.com | -0.150518354 | 0.123734456 |
| suitsupply.com | 0.088483884 | 0.112828866 |
| summitracing.com | -0.025209234 | 0.274640215 |
| sunbasket.com | -0.007993242 | 0.115915676 |
| sundancecatalog.com | -0.048805326 | 0.124185395 |
| superbiiz.com | 0.147637801 | 0.395308372 |
| superbrightleds.com | 0.014168936 | 0.283283647 |
| supply.com | 0.251075137 | 0.265052596 |
| supremenewyork.com | 0.206354279 | 0.291559999 |
| surfboard.com | 0.103535524 | 0.367575098 |
| surfstitch.com | 0.198584359 | 0.227691202 |
| surlatable.com | -0.001146498 | 0.192672749 |
| swansonvitamins.com | 0.025017835 | 0.264382715 |
| swap.com | -0.104962297 | 0.212259083 |
| sweetwater.com | 0.002786451 | 0.41831306 |
| swellbottle.com | 0.110232689 | 0.314838358 |
| swimoutlet.com | -0.167172302 | 0.223775654 |
| swisscolony.com | -0.157053151 | 0.259406926 |
| sylvane.com | 0.177376669 | 0.515044547 |
| tackledirect.com | -0.110521577 | 0.34869972 |
| tacklewarehouse.com | -0.029240984 | 0.144012922 |
| tacticalgear.com | -0.077331638 | 0.270015547 |
| tactics.com | 0.11320265 | 0.244705359 |
| talbots.com | -0.115707924 | 0.14844317 |
| tanga.com | -0.085949811 | 0.404346152 |
| target.com | -0.015754404 | 0.389473466 |
| targetsportsusa.com | 0.040636688 | 0.305652381 |
| tartecosmetics.com | -0.105170492 | 0.190052899 |
| teacollection.com | -0.051523796 | 0.104183546 |
| teamexpress.com | 0.091371312 | 0.323257119 |
| techforless.com | 0.01554915 | 0.367000675 |
| techrabbit.com | 0.058955454 | 0.316182866 |
| teespring.com | 0.095754948 | 0.298528793 |
| tempurpedic.com | 0.089513002 | 0.183435644 |
| tennis-warehouse.com | 0.098566629 | 0.190482355 |
| textbookrush.com | 0.194839567 | 0.209113024 |
| textbooks.com | 0.234611546 | 0.386755454 |
| the-house.com | 0.115746106 | 0.246576579 |
| thebay.com | 0.101340026 | 0.229677152 |
| theclymb.com | -0.019800433 | 0.178231281 |
| thecompanystore.com | 0.021954665 | 0.279695211 |
| thecubicle.us | 0.187888032 | 0.110009206 |
| thefryecompany.com | 0.093993912 | 0.157685404 |
| thegreatcourses.com | 0.007139709 | 0.227792022 |
| thehardwarehut.com | 0.099388908 | 0.359909509 |
| thenorthface.com | 0.079432513 | 0.163030607 |
| therealreal.com | 0.023414451 | 0.155466361 |
| thereformation.com | 0.220307528 | 0.090079042 |
| theshoppingchannel.com | 0.074851726 | 0.542547556 |
| thesource.ca | 0.107971381 | 0.34388224 |
| thetileapp.com | 0.080912666 | 0.219499862 |
| thewatchbox.com | 0.102286047 | 0.236263135 |
| thingsremembered.com | 0.031355043 | 0.416401175 |
| thirdlove.com | -0.144837337 | 0.228533517 |
| threadless.com | 0.113034628 | 0.240603438 |
| thredup.com | -0.141616446 | 0.312117959 |
| thriftbooks.com | 0.003406647 | 0.221122917 |
| thrivemarket.com | 0.057347317 | 0.342749738 |
| ties.com | 0.14767159 | 0.448341805 |
| tiffany.com | 0.055831894 | 0.149174408 |
| tileshop.com | 0.021184648 | 0.250308258 |
| tillys.com | 0.035948253 | 0.179562044 |
| timbuk2.com | 0.114661599 | 0.183107078 |
| timex.com | 0.01403905 | 0.218854404 |
| tipsyelves.com | 0.127230855 | 0.250042053 |
| tirebuyer.com | 0.042295453 | 0.377593718 |
| tirerack.com | -0.075569 | 0.206072339 |
| titlenine.com | -0.065571627 | 0.156570692 |
| tobi.com | 0.191481094 | 0.125258112 |
| tombihn.com | 0.094675612 | 0.215064166 |
| tomford.com | 0.153998141 | 0.223404591 |
| toms.com | -0.043401591 | 0.153749073 |
| tomtop.com | 0.008203285 | 0.396497987 |
| tooltopia.com | 0.090160956 | 0.359428113 |
| toolup.com | 0.123319902 | 0.631466042 |
| topshop.com | 0.145314952 | 0.228359717 |
| tortugabackpacks.com | 0.25251628 | 0.455282518 |
| toryburch.com | -0.028280487 | 0.251359585 |
| touchofclass.com | -0.027942949 | 0.334808304 |
| touchofmodern.com | -0.166227739 | 0.238012417 |
| towerhobbies.com | -0.024257648 | 0.24936245 |
| toysrus.com | 0.111879014 | 0.358813116 |
| tractorsupply.com | -0.094940294 | 0.208576945 |
| traegergrills.com | 0.017396747 | 0.378781459 |
| traetelo.com | 0.02012883 | 0.491461276 |
| traxxas.com | -0.06735538 | 0.277396697 |
| treehut.co | 0.02218356 | 0.357012545 |
| trekbikes.com | 0.027106187 | 0.19113816 |
| trinaturk.com | 0.052678224 | 0.30021756 |
| troybilt.com | -0.055225802 | 0.279222921 |
| truevalue.com | -0.024074491 | 0.298720561 |
| tsc.ca | 0.067573996 | 0.294733652 |
| tuftandneedle.com | 0.093901808 | 0.267500941 |
| turners.com | -0.182596618 | 0.155773109 |
| turntablelab.com | 0.198604101 | 0.348393845 |
| turtlebeach.com | 0.141186463 | 0.362022174 |
| ugg.com | 0.072868184 | 0.164947039 |
| uline.com | 0.107370046 | 0.227483903 |
| ullapopken.com | -0.138193316 | 0.223768262 |
| ulta.com | -0.028226261 | 0.211302543 |
| ultimateears.com | 0.113056386 | 0.310903163 |
| ultrasabers.com | -0.10815809 | 0.191383017 |
| unbeatablesale.com | 0.139626291 | 0.312839711 |
| uncommongoods.com | 0.060892205 | 0.283990984 |
| underarmour.com | 0.018525695 | 0.210337922 |
| uniformadvantage.com | -0.071818473 | 0.161458878 |
| uniqlo.com | 0.112958753 | 0.1194935 |
| unique-vintage.com | 0.027563937 | 0.182155022 |
| uniquesquared.com | 0.358268134 | 0.164417048 |
| uniquevintage.com | 0.252956704 | 0.99036123 |
| untilgone.com | -0.018889294 | 0.390666653 |
| untuckit.com | 0.011712524 | 0.231874216 |
| urbanoutfitters.com | 0.175498156 | 0.132006859 |
| us-appliance.com | 0.140911748 | 0.410056108 |
| us-mattress.com | 0.128040634 | 0.455953947 |
| us.humankinetics.com | 0.278216875 | 0.263788716 |
| usa.philips.com | 0.116992862 | 0.189684866 |
| uscargocontrol.com | 0.112305587 | 0.479855225 |
| uscutter.com | 0.125882648 | 0.30215826 |
| uspoloassn.com | 0.075337377 | 0.12131104 |
| v2.com | 0.101814665 | 0.506536334 |
| valuecityfurniture.com | -0.07552261 | 0.168008296 |
| vapes.com | -0.011746575 | 0.345412667 |
| vapewild.com | -0.153061665 | 0.222002267 |
| vapordna.com | -0.065169449 | 0.296494132 |
| varidesk.com | 0.20477112 | 0.274191263 |
| verabradley.com | -0.109645962 | 0.11791499 |
| vermontcountrystore.com | -0.083676063 | 0.541343195 |
| vermontteddybear.com | -0.021775088 | 0.257953978 |
| victoriassecret.com | -0.120433605 | 0.178116872 |
| vintageking.com | 0.35074741 | 0.304813874 |
| vintagetub.com | 0.086218828 | 0.326144498 |
| vipoutlet.com | 0.002228995 | 0.310279087 |
| viralstyle.com | -0.025972521 | 0.432772705 |
| vistaprint.com | 0.016882883 | 0.039603139 |
| vitaminshoppe.com | -0.04933265 | 0.433040347 |
| vitamix.com | 0.079700723 | 0.253456547 |
| vivino.com | 0.194306744 | 0.500030309 |
| vizio.com | 0.016285857 | 0.380084223 |
| vminnovations.com | 0.067469952 | 0.472447913 |
| walgreens.com | -0.102299665 | 0.186697437 |
| walmart.com | -0.033997644 | 0.246196884 |
| warbyparker.com | 0.054707228 | 0.176524302 |
| waterfilters.net | 0.108739155 | 0.481199519 |
| wayfair.com | -0.037134924 | 0.294045558 |
| wbmason.com | 0.083175096 | 0.004272289 |
| weathertech.com | -0.087161732 | 0.169003746 |
| weber.com | 0.02603137 | 0.361256336 |
| webeyecare.com | 0.0120778 | 0.147107885 |
| webyshops.com | 0.236880339 | 0.646124317 |
| weightwatchers.com | -0.012757361 | 0.154816995 |
| well.ca | 0.228048404 | 0.328673405 |
| westmarine.com | 0.000479439 | 0.355962797 |
| whiteflowerfarm.com | 0.021715092 | 0.326290714 |
| wholelattelove.com | 0.105851781 | 0.299058005 |
| williams-sonoma.com | 0.04411063 | 0.277056011 |
| windsorstore.com | 0.002047035 | 0.112689251 |
| windupwatchshop.com | 0.022599692 | 0.264033309 |
| wine.com | 0.149336093 | 0.417810154 |
| winechateau.com | 0.14973079 | 0.365034022 |
| wineenthusiast.com | 0.030065354 | 0.443086752 |
| winelibrary.com | 0.188964593 | 0.459121701 |
| woodcraft.com | 0.03824739 | 0.328405912 |
| woodlanddirect.com | 0.056548801 | 0.27948834 |
| workingperson.com | 0.079809676 | 0.44081994 |
| yaleappliance.com | 0.170107798 | 0.517477744 |
| yamibuy.com | 0.231443097 | 0.067984341 |
| yandy.com | -0.097956205 | 0.240661447 |
| yeti.com | 0.024579932 | 0.160619749 |
| ylighting.com | 0.245396817 | 0.323752697 |
| yugster.com | 0.222781587 | 0.495818176 |
| zagg.com | -0.03180472 | 0.181420883 |
| zara.com | -0.018817859 | 0.250665031 |
| zavvi.com | 0.065412606 | 0.396451233 |
| zazzle.com | 0.122961894 | 0.260827183 |
| zennioptical.com | -0.073692648 | 0.13685979 |
| zgallerie.com | -0.0578888 | 0.153143905 |
| zmodo.com | 0.059667672 | 0.225644049 |
| zola.com | -0.00789852 | 0.289982285 |
| zumiez.com | 0.016381999 | 0.229024725 |
| zzounds.com | -0.033101382 | 0.276615314 |